\documentclass[twocolumn,aps,superscriptaddress,showpacs,nofootinbib,floatfix]{revtex4-1}
\usepackage{epsfig,bm,feynmf}
\usepackage{graphics}
\usepackage{amsmath}
\usepackage[normalem]{ulem}  
\usepackage[dvips]{color} 

\renewcommand\sout{\bgroup \color{red} \ULdepth=-.5ex \ULset}

\begin{document}


\title{Chiral symmetry restoration versus deconfinement in heavy-ion collisions at high baryon density}

\author{W. Cassing}
\affiliation{Institut f\"{u}r Theoretische Physik, Universit\"{a}t Gie$\beta$en, Germany}

\author{A. Palmese}
\affiliation{Institut f\"{u}r Theoretische Physik, Universit\"{a}t Gie$\beta$en, Germany}

\author{P. Moreau}
\affiliation{Frankfurt Institute for Advanced Studies, Johann Wolfgang Goethe Universit\"{a}t, Frankfurt am Main, Germany}
\affiliation{Institute for Theoretical Physics, Johann Wolfgang Goethe Universit\"{a}t, Frankfurt am Main, Germany}

\author{E. L. Bratkovskaya}
\affiliation{Frankfurt Institute for Advanced Studies, Johann Wolfgang Goethe Universit\"{a}t, Frankfurt am Main, Germany}
\affiliation{Institute for Theoretical Physics, Johann Wolfgang Goethe Universit\"{a}t, Frankfurt am Main, Germany}


\begin{abstract}
We study the production of strange hadrons in nucleus-nucleus
collisions from 4 to 160 A GeV within  the Parton-Hadron-String
Dynamics (PHSD) transport approach that is extended to incorporate
essentials aspects of chiral symmetry restoration (CSR) in the
hadronic sector (via the Schwinger mechanism) on top of the
deconfinement phase transition as implemented in PHSD. Especially
the $K^+/\pi^+$ and the $(\Lambda+\Sigma^0)/\pi^-$ ratios in central
Au+Au collisions are found to provide information on the relative
importance of both transitions. The modelling of chiral symmetry
restoration is driven by the pion-nucleon $\Sigma$-term in the
computation of the quark scalar condensate $<q {\bar q}>$ that
serves as an order parameter for CSR and also scales approximately
with the effective quark masses $m_s$ and $m_q$. Furthermore, the
nucleon scalar density $\rho_s$, which also enters the computation
of $<q {\bar q}>$, is evaluated within the nonlinear $\sigma-\omega$
model which is constraint by Dirac-Brueckner calculations and low
energy heavy-ion reactions. The Schwinger mechanism (for string
decay) fixes the ratio of strange to light quark production in the
hadronic medium.  We find that above $\sim$80 A GeV the reaction
dynamics of heavy nuclei is dominantly driven by partonic
degrees-of-freedom such that traces of the chiral symmetry
restoration are hard to identify. Our studies support the conjecture
of 'quarkyonic matter' in heavy-ion collisions from about 5 to 40 A
GeV and provide a microscopic explanation for the maximum in the
$K^+/\pi^+$ ratio at about 30 A GeV which only shows up if a
transition to partonic degrees-of-freedom is incorporated in the
reaction dynamics and is discarded in the traditional hadron-string
models.

\end{abstract}

\pacs{25.75.Nq, 25.75.Ld, 25.75.-q, 24.85.+p, 12.38.Mh}
\keywords{}

\maketitle

\section{Introduction}
According to  Quantum Chromo Dynamics (QCD)
\cite{lQCD,lqcd0,LQCDx,Peter,Lat1,Lat2}, matter changes its phase at
high temperature and density and bound (colorless) hadrons dissolve
to interacting (colored) quarks and gluons in the Quark-Gluon-Plasma
(QGP). Along with this deconfinement phase transition at low quark
chemical potential $\mu_q$ a restoration of chiral symmetry (CSR) is
observed in lattice QCD (lQCD) calculations at roughly the same
critical temperature or energy density. Since at low $\mu_q$ the
phase change is a crossover both transitions do not (have to) occur
at the same temperature. The study of the phase boundaries and the
properties of the QGP are the main goal of several present and
future heavy-ion experiments at SPS (Super Proton Synchrotron), RHIC
(Relativistic Heavy-Ion Collider), LHC (Large Hadron Collider) and
the future FAIR (Facility for Antiproton and Ion Research) as well
as NICA (Nuclotron-based Ion Collider  fAcility) \cite{QM2014}.
Since the QGP is created only for a short time (of a couple of fm/c)
it is quite demanding to study its properties and to find sensible
probes for chiral symmetry restoration as well as for the
deconfinement transition. Only by the measurement of the 'bulk'
light hadrons, electromagnetic probes (dileptons and photons), heavy
mesons, jets and related correlations we might be able to
disentangle the different physics at the phase boundaries especially
at high quark chemical potential $\mu_q$ where a first order
transition might take place \cite{CBMbook,Ch1,Ch2,BJS}.

The question of chiral symmetry restoration at high baryon density
and/or high temperature is of fundamental interest in itself and
dilepton studies in nucleus-nucleus collisions have been driven by
the notion to find signatures for a CSR. As noted above  the
situation is less clear for finite baryon density where QCD sum rule
studies indicate a linear decrease of the scalar quark condensate
$<\bar q q>$ -- which is nonvanishing in the vacuum due to a
spontaneous breaking of chiral symmetry -- with baryon density
$\rho_B$ towards a chiral symmetric phase characterized by $<\bar{q}
q> \approx$  0 \cite{Weise,Birse}. This decrease of the scalar quark
condensate is expected to lead to a change of the hadron properties
with density and temperature, i.e. in a chirally restored phase the
vector and axial vector currents should become equal
\cite{Kochr,Zahed,GEB,GEB1,Koch}; the latter implies that e.g. the
$\rho$ and $a_1$ spectral functions should become identical. Since
the scalar quark  condensate $<q\bar{q}>$ is not a direct
observable, its manifestations should be found indirectly in
different hadronic abundancies and spectra or particle ratios like
$K^+/\pi^+$, $(\Lambda+\Sigma^0)/\pi^-$ etc.

Nowadays, our knowledge about the hadron properties at high
temperature or baryon density is based on heavy-ion experiments
from SchwerIonen-Synchrotron (SIS) to SPS energies where hot and dense nuclear
systems are produced on a timescale of a few fm/c together with partonic
subsystems. The information from ultra-relativistic nucleus-nucleus
collisions at RHIC or LHC essentially addresses
the dominant partonic phases at low quark chemical potential $\mu_q$, a region
which can be well addressed by lQCD, too. This knowledge, however, does not allow
for proper extrapolations to the properties of QCD at high baryon density
which has been the major motivation for the future construction of the FAIR/NICA
facilities and is in the focus of the Beam-Energy-Scan (BES) program at RHIC.
However, any informations on the properties of hadrons in the nuclear
environment are obtained from the comparison of experimental data with nonequilibrium kinetic transport theory
\cite{Stoecker,Bertsch,Cass90,Hart,HSD,Koreview,UrQMD1,Brat2012,Marty}.  As a genuine
feature of transport theories in the hadronic sector there are two essential ingredients:
i.e. the {\it baryon (and meson) scalar } and {\it vector
self-energies} as well as {\it in-medium elastic} and {\it
inelastic cross sections} (or transition matrix elements) for all hadrons involved. Whereas in the
low-energy regime these 'transport coefficients' can be calculated
in the Dirac-Brueckner approach starting from the bare
nucleon-nucleon interaction \cite{Bot,Mal,T3,Laura,EX4} this is no longer
possible at high baryon density $(\rho_B \geq 2$-$3 \rho_0)$ and
high temperature, since the number of independent hadronic degrees-of-freedom
increases drastically and the interacting hadronic
system should approach a phase with $<q\bar{q}> \approx$ 0
\cite{Weise,Kochr,GEB,GEB1} as mentioned before.

In this work we will concentrate on excitation functions of hadronic
observables from SIS to SPS energies with the aim to find out
optimal experimental conditions to search for 'traditional'
phenomena such as strangeness enhancement in nucleus-nucleus
collisions \cite{rafelski,stock} or the 'horn' in the $K^+/\pi^+$
ratio \cite{GG99,GGS11}. Both phenomena have been addressed to a
deconfinement transition. Indeed,  the actual
experimental observation could not be described within conventional
hadronic transport theory \cite{Jgeiss,Brat04,Weber} and remained a major
challenge for microscopic approaches. Only within hybrid approaches
\cite{Petersen} or three-fluid hydrodynamics \cite{hydro2} a
description of the 'horn phenomenon' could be achieved due to the
assumption of chemical equilibrium in the hydro phase. This also
holds true for the statistical model fits assuming full chemical
equilibration \cite{ABS06}. However, hadronic interaction rates
showed up to be too slow to reach chemical equilibrium in these
nucleus-nucleus collisions \cite{Brat00}. This was
also found for the partonic stage in Ref. \cite{Vitalii}. Accordingly, the
quest for a microscopic explanation of the $K^+/\pi^+$ 'horn'
remained.

In 2006 McLerran and Pisarski suggested that a new form of matter
might exist at high baryon chemical potential \cite{Rob1} where the
degrees-of-freedom are still confined but chiral symmetry is
restored such that parity doublets appear in the excitation
spectrum. This lead to an extended scenario for the phase diagram of
strongly interacting matter \cite{Rob2,sasaki}. Although a clear
separated phase of such 'quarkyonic matter' might be overtaken one
could expect a partially chiral restored phase before deconfinement
sets in as suggested also by Nambu-Jona-Lasinio models
\cite{Nambu1961,Klevansky1992,NJL,Juan}.

Our studies are performed within the PHSD transport approach that
has been described in Refs. \cite{PHSD,PHSDrhic}. PHSD incorporates
explicit partonic degrees-of-freedom in terms of strongly
interacting quasiparticles (quarks and gluons) in line with an
equation-of-state from lattice QCD (lQCD) as well as dynamical
hadronization and hadronic elastic and inelastic collisions in the
final reaction phase. This approach has been tested for $p+p$, $p+A$
and $A+A$ collisions from the SPS to LHC energy regime
\cite{PHSD,PHSDrhic,Volo,Linnyk}. We recall that longitudinal and
transverse momentum spectra of nucleons, pions, kaons and
antibaryons have been successfully reproduced within PHSD within the
full energy range from the upper SPS to LHC energies as well
dilepton and photon observables. Moreover, the collective flow
coefficients $v_n$ for the azimuthal angular distributions were
found to be well in line with the PHSD calculations as well as the
suppression of hard probes such as charm quarks at RHIC energies
\cite{TSong}. However, in all these studies the question of chiral
symmetry restoration has been discarded since the observables
analyzed were driven by the dominant deconfinement transition and
the parton dynamics in the QGP. Only when going down in bombarding
energy to the Alternating-Gradient-Synchrotron (AGS) regime severe
discrepancies were found \cite{Volo14} in the directed proton, pion
and kaon flows as well as in the  strangeness production. The actual
results at AGS energies are close to those from the Hadron-String
Dynamics (HSD) transport approach (without a partonic phase) in
Refs. \cite{Brat04,Weber,BratPRL}. This does not come as a surprise
since at low energy densities ($\epsilon < 0.5 $ GeV/fm$^3$) the
PHSD merges with HSD. By observing that the discrepancies show up
already within HSD - and are very similar in PHSD - we conclude that
the deconfinement phase transition does not show 'responsibility'
for these discrepancies.

We here suggest that the missing 'strangeness' is due to the neglect
of chiral symmetry restoration (in the Schwinger mechanism of string
decay) at high baryon density. Accordingly, a partial restoration of
chiral symmetry may be achieved in the hadronic but confined phase.
Such a situation might be attributed to 'quarkyonic matter' out-off
equilibrium. To investigate this proposal we extend the existing
PHSD transport approach \cite{PHSD}  to include essential facets of
chiral symmetry restoration in terms of the Schwinger mechanism for
string decay. Since PHSD has been successfully applied to describe
the final distribution of mesons (with light quark content) from SPS
up to LHC energies \cite{PHSD,PHSDrhic,Volo,Linnyk}, it provides a
solid framework for the description of the creation, expansion and
hadronization of the QGP as well as the hadronic expansion with
partly resonant scatterings (like $\pi + N \leftrightarrow \Delta$
or $\pi + \pi \leftrightarrow \rho$).

This study is organized as follows:
In Sec. \ref{PHSD} we recall the basic ideas of the PHSD approach and describe the
evaluation of the scalar quark condensate as well as the modified string decay in line with CSR.  Sec. III
is devoted to the actual computation of particle spectra and particle ratios in central Au+Au
collisions from 4 to 158 A GeV in different limits, i.e. with and without CSR, with and without
partonic degrees-of-freedom. A summary completes this work in Sec.~\ref{summary}.

\section{The PHSD transport approach}\label{PHSD}

\subsection{Basic concepts}
The Parton-Hadron-String Dynamics (PHSD) transport
approach~\cite{PHSD,PHSDrhic} is a microscopic covariant dynamical
model for strongly interacting systems formulated on the basis of
Kadanoff-Baym equations \cite{Kadanoff1,Kadanoff2,Cassing:2008nn}
for Green's functions in phase-space representation (in first order
gradient expansion beyond the quasiparticle approximation). The
approach consistently describes the full evolution of a relativistic
heavy-ion collision from the initial hard scatterings and string
formation through the dynamical deconfinement phase transition to
the strongly-interacting quark-gluon plasma (sQGP) as well as
hadronization and the subsequent interactions in the expanding
hadronic phase as in the Hadron-String-Dynamics (HSD) transport
approach \cite{HSD}. The transport theoretical description of quarks
and gluons in the PHSD is based on the Dynamical Quasi-Particle
Model (DQPM) for partons that is constructed to reproduce lQCD
results for a quark-gluon plasma in thermodynamic
equilibrium~\cite{Cassing:2008nn} on the basis of effective
propagators for quarks and gluons. The DQPM is thermodynamically
consistent and the effective parton propagators incorporate finite
masses (scalar mean-fields) for gluons/quarks as well as a finite
width that describes the medium dependent reaction rate. For fixed
thermodynamic temperaure $T$ the partonic width's $\Gamma_i(T)$ fix
the effective two-body interactions that are presently implemented
in the PHSD~\cite{Vitalii}. The PHSD differs from conventional
Boltzmann approaches in a couple of essential aspects:
\begin{itemize}
\item{it incorporates dynamical quasi-particles due to the finite
width of the spectral functions (imaginary part of the propagators)
in line with complex retarded selfenergies;} \item{it involves
scalar mean-fields that substantially drive the collective flow in
the partonic phase;} \item{it is based on a realistic equation of
state from lattice QCD and thus reproduces the speed of sound
$c_s(T)$ reliably;} \item{the hadronization is described by the
fusion of off-shell partons to off-shell hadronic states (resonances
or strings);} \item{all conservation laws (energy-momentum, flavor
currents etc.) are fulfilled in the hadronization contrary to
coalescence models;} \item{the effective partonic cross sections no
longer are given by pQCD and are 'defined' by the DQPM in a
consistent fashion. These cross sections are probed by transport coefficients
(correlators) in thermodynamic equilibrium by performing PHSD
calculations in a finite box with periodic boundary conditions
(shear- and bulk viscosity, electric conductivity, magnetic
susceptibility etc. \cite{Vitaly2,Ca13}).}
\end{itemize}

An actual nucleus-nucleus collision within PHSD proceeds as follows:
in the beginning of a relativistic heavy-ion collisions
color-neutral strings (described by the FRITIOF LUND
model~\cite{FRITIOF}) (including PYTHIA 6.4 \cite{Sjostrand:2006za})
are produced in hard scatterings of nucleons
from the impinging nuclei. These strings are dissolved into
'pre-hadrons' with a formation time of $\sim$ 0.8 fm/c in their rest
frame, except {for} the 'leading hadrons', i.e. the fastest residues
of the string ends, which can re-interact (practically instantly)
with hadrons with a reduced cross sections in line with quark
counting rules. If, however, the local energy density is larger than
the critical value for the phase transition, which is taken to be
$\sim$ 0.5 ${\rm GeV/ fm^3}$ in line with lQCD \cite{LQCDx}, the
pre-hadrons melt into (colored) effective quarks and antiquarks in
their self-generated repulsive mean-field as defined by the
DQPM~\cite{Cassing:2008nn}. In the DQPM the quarks, antiquarks and
gluons are dressed quasiparticles and have temperature-dependent
effective masses and widths which have been fitted to  lattice
thermal quantities such as energy density, pressure and entropy
density. The nonzero width of the quasiparticles implies the
off-shellness of partons, which is taken into account in the
scattering and propagation of partons in the QGP on the same footing
(i.e. propagators and couplings).

The transition from the partonic to hadronic degrees-of-freedom (for
light quarks/antiquarks) is described by covariant transition rates
for the fusion of quark-antiquark pairs to mesonic resonances or
three quarks (antiquarks) to baryonic states, i.e. by the dynamical
hadronization. We already mention here that this hadronization
process is restricted to 'bulk' transverse momenta $p_T$  up to
$\sim $ 2 GeV and has to be replaced by fragmentation for high $p_T$
\cite{TSong} in future. Note that due to the off-shell nature of both partons
and hadrons, the hadronization process described above obeys all
conservation laws (i.e. four-momentum conservation and flavor
current conservation) in each event, the detailed balance relations
and the increase in the total entropy $S$.

In the hadronic phase
PHSD is equivalent to the hadron-strings dynamics (HSD) model
\cite{HSD} that has been employed in the past from SIS to SPS
energies. On the other hand the PHSD approach has been tested for
p+p, p+A and relativistic heavy-ion collisions from lower SPS to LHC
energies and been successful in describing a large number of
experimental data including single-particle spectra, collective flow ~\cite{PHSD,PHSDrhic,PHSDLHC,Volo}
as well as electromagnetic probes \cite{Linnyk} or charm observables \cite{TSong}.

\subsection{Strings in (P)HSD}
In the PHSD/HSD the string excitation and decay plays a decisive role for elastic and inelastic
collisions which has been well tested for hadronic reactions in vacuum in a wide energy range. We recall that in
the hadronic phase the high energy inelastic hadron-hadron
collisions are described by the FRITIOF model \cite{FRITIOF}, where
two incoming nucleons emerge the reaction as two excited color
singlet states, i.e. 'strings'. According to the FRITIOF model
a string is characterized by the leading constituent
quarks of the incoming hadron and a tube of color flux is supposed
to be formed connecting the rapidly receding string-ends. In the
PHSD approach baryonic ($qq-q$) and mesonic ($q-\bar{q}$) strings
are considered with different flavors ($q = u,d,s$). In the
uniform color field of the strings virtual $q\bar{q}$ or
$qq\bar{q}\bar{q}$ pairs are produced causing the tube to fission
and thus to create mesons or baryon-antibaryon pairs (with a formation time $\sim$ 0.8 fm/c). The
production probability $P$ of massive $s\bar{s}$ or
$qq\bar{q}\bar{q}$ pairs is suppressed in comparison to light
flavor production ($u\bar{u}$, $d\bar{d}$) according to a
Schwinger-like formula \cite{Schwinger}, i.e.
\begin{eqnarray}
{P(s\bar{s}) \over P(u\bar{u})} ={P(s\bar{s}) \over P(d\bar{d})} = \gamma_s = \exp\left(-\pi
{m_s^2-m_q^2\over 2\kappa} \right) ,
\label{schwinger}
\end{eqnarray}
with $\kappa\approx 0.176$~GeV$^2$ denoting the string tension
and $m_s, m_q=m_u=m_d$ the appropriate strange and light quark masses. Thus in
the Lund string picture the production of strangeness and
baryon-antibaryon pairs is controlled by the constituent quark (and
diquark) masses. Inserting the constituent (dressed) quark masses $m_u \approx 0.33$~GeV
and $m_s \approx 0.5$ GeV in the vacuum a value of $\gamma_s \approx 0.3$ is
obtained from Eq. (1). While the strangeness production in proton-proton
collisions at SPS energies is reasonably well reproduced with this
value, the strangeness yield for p~+~Be collisions at AGS energies
is underestimated by roughly 30\% (cf. \cite{Jgeiss}). For that
reason the relative factors used in the PHSD/HSD model are
\cite{Jgeiss}
\begin{eqnarray}
u:d:s:uu = \left\{
\begin{array}{ll}
1:1:0.3:0.07 &, \mbox{at SPS to RHIC } \\ 1:1:0.4:0.07 &,
\mbox{at AGS energies} ,
\end{array}
\right. \label{HSD-supp}
\end{eqnarray}
with a linear transition of the strangeness suppression factor
$\gamma_s$ as a function of $\sqrt{s}$ in between. These settings
have been fixed in Ref. \cite{Jgeiss} for HSD in 1998 and kept since
then also for PHSD.

Additionally a fragmentation function
$f(x,m_t)$ has to be specified, which is the probability
distribution for hadrons with transverse mass $m_T$ to acquire the
energy-momentum fraction $x$ from the fragmenting string,
\begin{eqnarray}
f(x,m_T)\approx {1 \over x} (1-x)^a \exp\left(-bm_T^2/x  \right),
\end{eqnarray}
with $a=0.23$ and $b=0.34$~GeV$^{-2}$ \cite{Jgeiss}.
These settings for the string decay to hadrons have been found to match well experimental
observations for particle production in p+p and p+A reactions \cite{HSD}. In these reactions
the vacuum constituent (dressed) quark masses $m_s$ and $m_q$ are relevant that, however,  should be different
in the nuclear medium as noted above.
We mention that in HSD/PHSD we include antinucleon annihilation into several mesons
while taking into account also  the inverse processes of $N\bar{N}$
creation in multi-meson interactions by detailed
balance~\cite{Ca02}.

\subsection{Extensions in PHSD with respect to HSD2.3}
The modifications in PHSD3.3 with respect to the HSD version 2.3
from 2003 employed in the previous studies in Refs.
\cite{Brat04,Weber} are as follows:
\begin{itemize}
\item{The energy-density cut for hadronic interactions ($\epsilon_c \leq $ 1 GeV/fm$^3$) has been reduced to
$\epsilon_c \leq $ 0.5 GeV/fm$^3$ because the unquenched lattice QCD results from the BMW Collaboration in 2009 \cite{lqcd0} were converging to
the lower critical energy density $\epsilon_c \approx$ 0.5 GeV/fm$^3$ for the deconfinement phase transition. Accordingly, inelastic hadronic collisions in local cells with energy density   0.5 GeV/fm$^3 \leq \epsilon \leq$ 1 GeV/fm$^3$ no longer happen in the actual version which leads to a reduction of the meson multiplicity (dominantly pions and kaons) by about 10 to 15\%. Since kaons and pions are effected by about the same reduction factor their ratios practically do not change. }
\item{The inverse processes of $N\bar{N}$ creation in multi-meson interactions by detailed
balance~\cite{Ca02} are now included by default.}
\item{A couple of strangeness exchange reactions have been added in meson-baryon and baryon-baryon collisions
following Refs. \cite{Song1,Song2,Song3}.}
\item{The hadronic reaction channels are symmetric now with respect to $m+B \leftrightarrow {\bar m} + {\bar B}$. This has
a minor impact at the laboratory energies considered in this study.}
\end{itemize}
After reviewing the basic concepts of the PHSD approach we now come
to the modeling of chiral symmetry restoration in the PHSD.

\subsection{The scalar quark condensate}
As is well known the scalar quark condensate $<q\bar{q}>$ is viewed as an order
parameter for the restoration of chiral symmetry at high baryon
density and temperature, however, it is not a quantity that can directly be
determined by experiment. Nevertheless, some close links to low energy constants
and nuclear quantities can be employed to determine the scalar quark condensate.

A reasonable estimate for the quark scalar condensate in dynamical
calculations has been suggested by Friman et al. \cite{Toneev98},
\begin{equation}
\frac{<q\bar{q}>}{<q\bar{q}>_V} = 1 - \frac{\Sigma_\pi}{f_\pi^2
m_\pi^2}\rho_S - \sum\limits_h{\sigma_h \rho_S^h \over f_\pi^2
m_\pi^2}, \label{condens2} \end{equation} where $\sigma_h$ denotes
the $\sigma$-commutator of the relevant mesons $h$. Furthermore,
$<q\bar{q}>_V$ denotes the vacuum condensate, $\Sigma_\pi \approx$
45 MeV is the pion-nucleon $\Sigma$-term, $f_\pi$ and $m_\pi$ the
pion decay constant and pion mass, respectively. Since for low
densities the scalar density $\rho_S$ in (\ref{condens2}) may be
replaced by the baryon density $\rho_B$, the change in the scalar
quark condensate starts linearly with $\rho_B$ and is reduced by a
factor $\approx$ 1/3 at saturation density $\rho_0 \approx$ 1/6
fm$^{-3}$. Note, however, that the value of $\Sigma_{\pi} $ is not
so accurately known; a recent analysis points towards a larger value
of $\Sigma_{\pi} \approx 59$ MeV \cite{Meissner} while actual lQCD
results \cite{Sternbeck} suggest a slightly lower value.
Accordingly, our following calculations - based on $\Sigma_{\pi} =
45$ MeV - have to be taken with some care.

For pions and mesons made out of light quarks and antiquarks we use
$\sigma_h = m_\pi/2$ \cite{Sternbeck} whereas for mesons with a
strange (antistrange) quark we adopt $\sigma_h = m_\pi/4$ according
to the light quark content. Within the same spirit the $\sigma$-term
for hyperons is taken as 2/3 $\Sigma_{\pi} \approx$ 30 MeV while for
$\Xi$'s we use 1/3 $\Sigma_{\pi} \approx$ 15 MeV. The vacuum scalar
condensate
 $ <q\bar{q}>_V$ is fixed by the Gell-Mann-Oakes-Renner (GOR) relation
\cite{GOR,Cohen}
\begin{equation} f_\pi^2 m_\pi^2
= - \frac{1}{2} (m_u^0 + m_d^0) <q\bar{q}>_V \end{equation}
to $<q\bar{q}>_V \approx $ - 1.6 fm$^{-3}$ for the bare quark masses $m_u^0 = m_d^0 \approx$ 7 MeV.
The scalar density of mesons (of type $h$) in (\ref{condens2}) is given by ($x=({\bf r}, t)$)
\begin{equation}
\label{mesons} \rho_S^h(x) = \frac{(2s+1) (2t+1)}{(2\pi)^3} \int d^3
{\bf p} \frac{m_h}{\sqrt{{\bf p}^2 + m_h^2}} f_h(x,{\bf
p}),
\end{equation}
with $f_h(x,{\bf p})$ denoting the meson phase-space distribution of
species $h$. In (\ref{mesons}) $s,t$ denote the discrete spin and
isospin quantum numbers, respectively. The last quantity in the
relation (\ref{condens2}), that still has to be determined for an
evaluation of the quark condensate, is the nucleon scalar density
$\rho_S$.

\subsection{The nuclear scalar density}
The scalar density of nucleons $\rho_S$ can be calculated in line
with (\ref{mesons}) by replacing the mass and momentum by the
effective quantities
\begin{equation} \label{mass} m_N^*(x)= m_N^v- g_s \sigma(x)  \end{equation}
with $m_N^v$ denoting the nucleon mass in vacuum. In Eq.
(\ref{mass}) the scalar field $\sigma(x)$ mediates the scalar
interaction with the surrounding medium while $g_s$ is a coupling.
When including self-interactions of the $\sigma$-field up to 4th
order \cite{Boguta} the scalar field is determined locally by the
nonlinear gap equation \cite{Boguta,Lang}
\begin{equation} \label{gap}
m_s^2 \sigma(x) + B \sigma^2(x) + C \sigma^3(x) = g_s \rho_S
\end{equation} $$=
 g_s d \int \frac{d^3 p}{(2 \pi)^3}
\frac{m_N^*(x)}{\sqrt{p^3+m_N^{*2}}} f_N(x,{\bf p})   $$ with $d=4$
in case of isospin symmetric nuclear matter. The parameters $g_s,
m_s, B, C$ are fixed in the non-linear $\sigma-\omega$ model for
nuclear matter \cite{Lang} and are displayed in Table I specifying
also the vector coupling $g_v$ and vector meson mass $m_v$. We will
use the NL3 set in the following with a compressibility $K = 380$
MeV and effective mass $m^*/m = 0.7$ at normal nuclear matter
density. In order to obtain an estimate on the uncertainties of our
results we have also used the sets NL1 and ML2. We recall that in
the non-linear $\sigma-\omega$ model the energy density for
symmetric nuclear matter is given by \cite{Lang}
\begin{equation} \label{eps}
\epsilon = U(\sigma) + \frac{g_v^2}{2 m_v^2} \rho_N^2 + d \int
\frac{d^3 p}{(2\pi)^3}
 \ E^*({\bf p}) \left(  N_{c}({\bf p})+  N_{d}({\bf p})\right)  , \end{equation}
with $$E^*({\bf p}) = \sqrt{{\bf p}^2 + m_B^{*2}},$$
$$ U(\sigma) =
\frac{m_s^2}{2} \sigma^2 +  \frac{B}{3} \sigma^3 +  \frac{C}{4} \sigma^4$$ while $\rho_N$
denotes the baryon density and $N_c({\bf p})$ and $N_{d}({\bf p})$
the particle/antiparticle numbers at fixed momentum ${\bf p}$.

\begin{table}
\vspace{1em}
\label{tab1}
\begin{tabular}{|c|c|c|c|} \hline
                      & NL1    &  ML2  &  NL3  \\ \hline
 $g_s$                &  6.91    &   9.28   & 9.50  \\ \hline
 $g_v$                &  7.54  &   10.59    & 10.95  \\ \hline
 $B$ (1/fm)           &  -40.6  &  5.1   & 1.589  \\ \hline
 $C$                  &   384.4 &   9.8   & 34.23  \\ \hline
 $m_s$ (1/fm)         &  2.79        &   2.79  & 2.79      \\ \hline
 $m_v$  (1/fm)        &  3.97        &  3.97  & 3.97      \\ \hline
 $K$ (MeV)            &  380         &   354  &  380       \\ \hline
 $m^*/m$              & 0.83         & 0.68 &   0.70       \\ \hline
\end{tabular}

\vspace{0.3cm} \caption{Parameter sets NL1, ML2 and NL3 for the
nonlinear $\sigma-\omega$ model employed in the transport
calculations \protect\cite{Lang}.}
\end{table}

\begin{figure}[thb]
\includegraphics[width=0.48\textwidth]{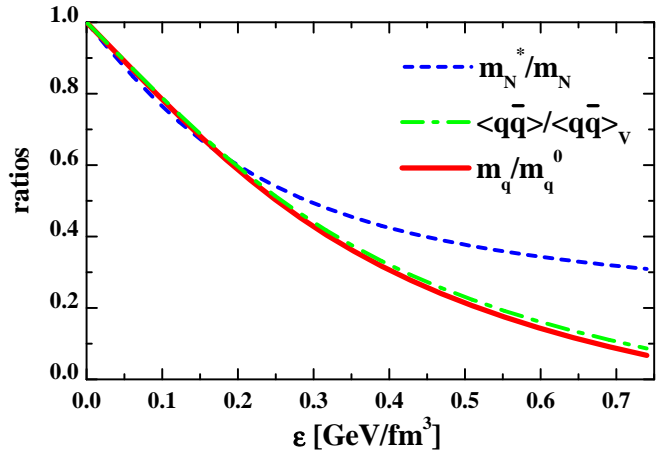}
\caption{The ratio $m^*_N/m_N$ as a function of the energy density
$\epsilon$ (\ref{eps}) at $T=0$ in the non-linear $\sigma- \omega$
model for the parameter-set  NL3 in comparison to the respective
ratio for the scalar quark condensate (solid red line) from
(\protect\ref{condens2}) (using $\Sigma_\pi $ = 45 MeV). The effective light
quark mass ratio $m_q/m_q^0$ (\ref{mus}) is displayed by the
dot-dashed green line for comparison.}
 \label{fig0}
\end{figure}

Actual results for the effective nucleon mass (divided by the vacuum
mass) at temperature $T=0$ are  displayed in Fig. \ref{fig0} by the
dashed (blue) line (for NL3) as a  function of the energy density
$\epsilon$ given by Eq. (\ref{eps}). We find an almost linear
decrease of the effective nucleon mass with energy density with a
slope that essentially scales with the effective mass $m^*_N/m_N^v$
at normal nuclear matter density $\rho_0$ (cf. Table I). Since for
$T=0$ there are no thermal mesons the resulting ratio for the scalar
quark condensate ${<q\bar{q}>}/{<q\bar{q}>_V}$ (\ref{condens2}) is
entirely fixed by the scalar nucleon density $\rho_S$ (for given
$\Sigma_\pi \approx$ 45 MeV). The resulting ratio is shown in Fig.
\ref{fig0} by the solid (red) line as a function of the energy
density $\epsilon$ while the ratio of the light quark mass to its
constituent mass $m_q/m_q^0$ is displayed by the dot-dashed (green)
line which practically coincides with the ratio of the scalar
condensate according to Eq. (\ref{mus}). We note in passing that
very similar results are obtained for the parameter set ML2 which
provides a very good fit to the Dirac-Brueckner results for the
nuclear equation of state from Refs. \cite{DBR,Dejong89} and is the
default parameter set used presently in the PHSD. We recall that the
non-linear $\sigma-\omega$ model has been often employed in the
description of heavy-ion reactions at SIS energies and its
parameters been determined in comparison to nuclear flow data
\cite{Lang}.

The question now arises, if there are proper experimental
observables that are especially sensitive to the variation of the quark scalar
condensate in a dense and hot medium. It has been suggested in Ref. \cite{Weber}
that when gating on central collisions of Au~+~Au (or
Pb~+~Pb) such phenomena should show up in the excitation
functions of suitable observables. We note that in a pure hadronic
transport approach we expect a smooth behaviour of practically all observables with
bombarding energy due to an increase of thermal excitation energy \cite{CBJ00}.
This is not so obvious for the PHSD approach where a gradual
transition from hadronic excitations to strings and to partonic
degrees-of-freedom is involved. We will argue that the strangeness ratios $K^+/\pi^+$, $K^-/\pi^-$ etc.
are suitable candidates (see below).

\subsection{The string decay in a hot and dense medium}
We recall that for the bombarding energies of interest from 4 to 158 A GeV the dominant particle
production in nucleus-nucleus reactions proceeds via string formation and decay.
The formation and decay of strings in the vacuum has been
investigated by decades and is described by the Schwinger mechanism
of quark-antiquark pair production  \cite{Schwinger} as discussed in Section IIB. In the
Schwinger formula for the strangeness fraction $s/u$
(\ref{schwinger}) the string tension $\kappa$ is determined
experimentally (or by lQCD) as well as the effective masses $m_s,
m_q$ for the dressed quarks. In line with common understanding this
dressing is due to a scalar coupling to the vacuum condensate $<q
{\bar q}>_v$ which - according to the previous Subsection - vanishes
with increasing baryon density and/or temperature. In first order
the dressed quark masses are expected to scale with the ratio
(\ref{condens2}) as
\begin{equation}   \label{mss}
m_s^* = m_s^0 + (m_s^v-m_s^0)\left| \frac{<q\bar{q}>}{<q\bar{q}>_V} \right| , \end{equation}
\begin{equation}   \label{mus}
m_q^* = m_q^0 + (m_q^v-m_q^0) \left| \frac{<q\bar{q}>}{<q\bar{q}>_V} \right| ,
\end{equation} using $ m_s^0 \approx$ 100 MeV and $m_q^0 \approx 7$
MeV for the bare quark masses.
In the hadronic phase the ratio (\ref{schwinger})
increases with decreasing scalar quark condensate as long as the
string tension $\kappa$ remains approximately constant. In fact,
lQCD results for the string tension below $T_c$ show roughly a
constant value while the string tension rapidly drops above $T_c$
since no coherent electric color fields (strings) can be formed
anymore. Accordingly, Eq. (\ref{schwinger}) no longer applies for
the deconfined phase and the ratio $s/u$ in PHSD is  fixed to $\sim 1/3$
by comparison to the strangeness production at RHIC and LHC energies
observed experimentally.
  \begin{figure}[thb]
\includegraphics[width=0.48\textwidth]{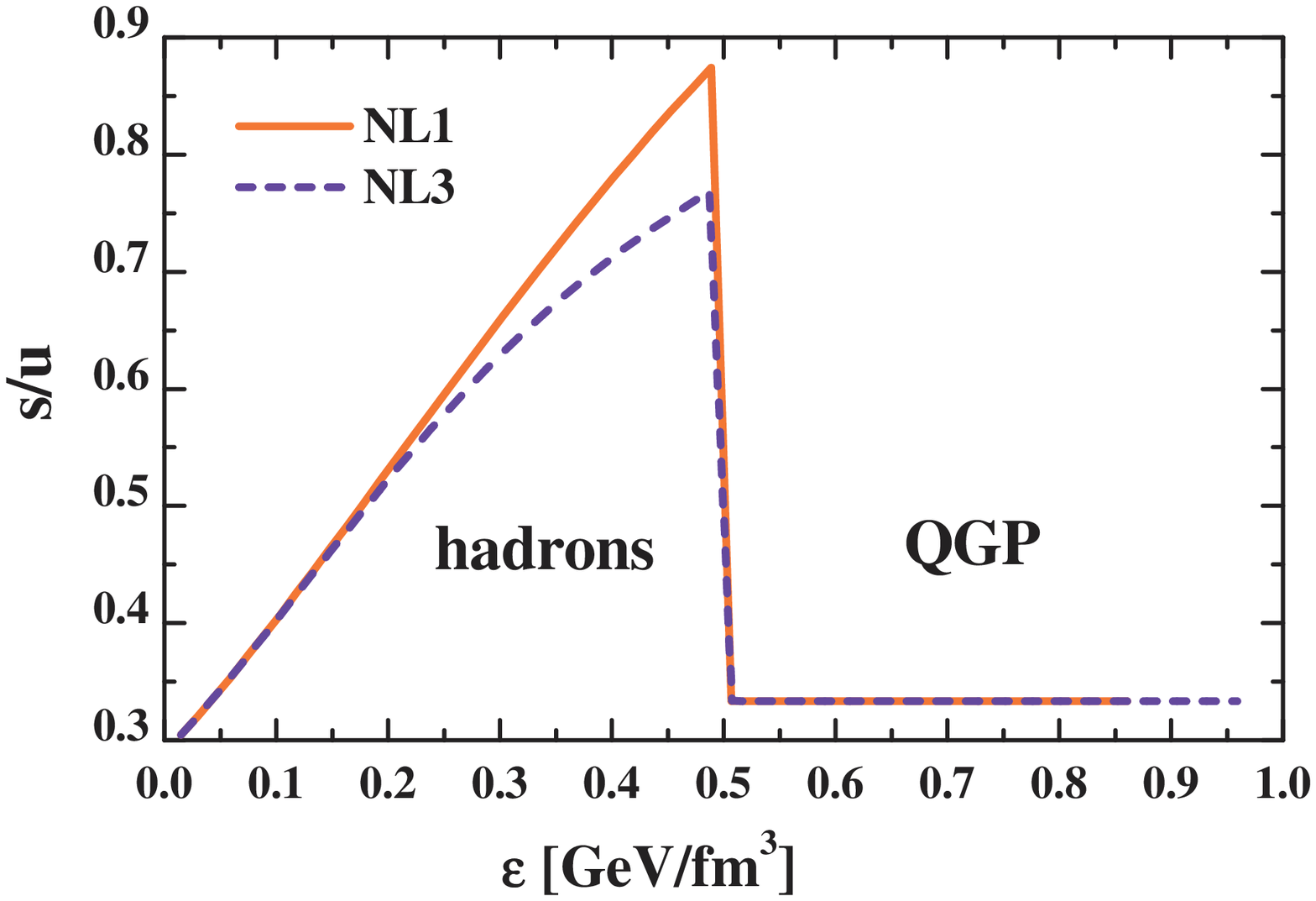}
\caption{The strangness ratio $s/u$ in the string decay
(\ref{schwinger}) as a function of the energy density $\epsilon$
(\ref{eps}) as evaluated within the non-linear $\sigma-\omega$ model
for the parameter sets NL3 and NL1  for $T=0$ on the basis of Eqs.
(\ref{mss}), (\ref{mus}) and (\ref{condens2}).}
 \label{fig1}
\end{figure}

In order to illustrate our main conjecture we show in Fig.
\ref{fig1} the ratio $s/u$ as a function of the energy density
$\epsilon$ at temperature $T$=0 as evaluated by Eq. (\ref{eps}). In
this case only nucleons contribute to the scalar density in Eq.
(\ref{condens2}) and the hadronic energy density which is (apart
from slight corrections due to the interaction energy) roughly given
by $\epsilon \approx m_N \rho_N$, where $m_N$ is the vacuum nucleon mass
and $\rho_N$ the nuclear density. It is seen that the ratio $s/u$
rises with nucleon density up to $\epsilon \approx$ 0.5 GeV/fm$^3$
and drops to a value of 1/3 in the deconfined phase in case of the
PHSD. The sensitivity to the nuclear model (NL1 versus NL3) is
moderate. The set ML2 practically gives the same results as the set
NL3. Accordingly, the approach to CSR occurs in the hadronic phase
and should be seen experimentally for local energy densities below
$\epsilon_c \approx$ 0.5 GeV/fm$^3$ since there is no more any
'string decay' in the partonic phase above $\epsilon_c$ due to the direct
conversion of energy and momentum to the massive quasiparticles of
the QGP.

In order to implement this scenario of 'chiral symmetry restoration'
in the PHSD code we solve the gap equation (\ref{gap}) for each cell
in space-time in order to determine the scalar nucleon density
$\rho_S$, the scalar quark condensate by Eq. (\ref{condens2}) and
the strangeness ratio by Eq. (\ref{schwinger}) which drives the
string decay in each local cell. Note that in the case of HSD
(without any deconfined partonic phase) the ratio $s/u$ increases
further with energy density $\epsilon$ up to the limiting values
given by bare masses $m_s=m_s^0$ and $m_q=m_q^0$ in Eq.
(\ref{schwinger}). Accordingly, one has to expect an overestimation
of strangeness production in the HSD when implementing CSR via
(\ref{schwinger}) at high bombarding energies where the scalar quark
condensate is vanishing in the overlap zone of the reaction.

We close this Section by noting that the PHSD approach has been
extended to include the essential features of chiral symmetry
restoration in the hadron production via the Schwinger mechanism
(\ref{schwinger}). One might criticise that the value of
$\Sigma_\pi$ as well as the nuclear equation of state (NL1, ML2,
NL3) are still uncertain to some extent but  these 'error bars' are small
compared to the leading order terms in the dynamics. Furthermore, a
fully consistent approach has to include the chiral partners of
$0^-$ and $1^-$ mesons, i.e. the (broad) scalar $0^+$ and axial
vector $1^+$ mesons. Also on the baryonic side the chiral partners
to the lightest baryons have to be incorporated in the transport
approach with dynamical spectral functions that become identical in
the chiral limit. Furthermore, the baryon propagators have to be refined
in including the real and imaginary parts of the scalar and vector selfenergies
also for the dynamical evolution of the system and the computation of in-medium
transition rates.  Since in principle the PHSD is suited for this
task due to its off-shell nature we delay a self-consistent
treatment of the chiral dynamics to a future more elaborate study.

\section{Results for central Au+Au collisions from 4 to 158 A GeV}\label{comparison}

\begin{figure}[thb]
\includegraphics[width=0.48\textwidth]{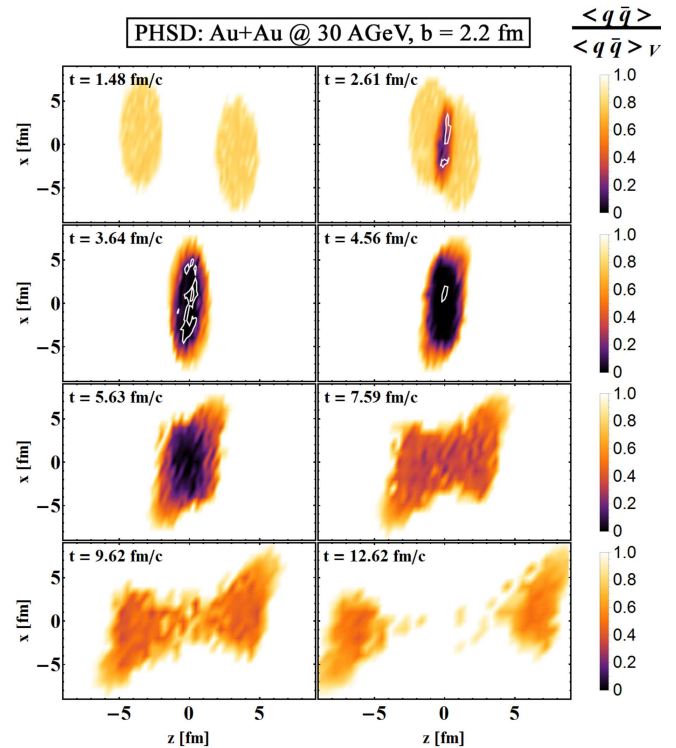}
\caption{The ratio (\ref{condens2}) for the scalar quark condensate
as a function of $x$ and $z$ (for $y=0$) at different times $t$ for
a central Au+Au collisions at 30 A GeV employing the parameter set
NL3 for the computation of the baryon scalar density. The white
borderline separates the space-time regions of deconfined matter to
hadronic matter. }
 \label{fig2}
\end{figure}

We recall that numerical results for the space-time dependence of
the scalar condensate ratio (\ref{condens2}) from HSD calculations
have been shown already in Figs. 3 and 4 of Ref. \cite{CBJ00} for
central Au~+~Au collisions at 6~A$\cdot$GeV and 20~A$\cdot$GeV,
respectively. For an illustration we present here additionally  the
ratio (\ref{condens2}) in Fig. \ref{fig2} as a function of $x$ and
$z$ (for $y=0$) at different times $t$ for a central Au+Au
collisions at 30 A GeV. Whereas in the approach phase the ratio
drops to about 2/3 inside the impinging nuclei (cf. Fig. 1) the scalar quark
condensate practically vanishes in the full overlap phase from about
3 to 7 fm/c and regains the vacuum value only in the late expansion
phase as noted before. However, for all times from contact ($\sim$ 2.5
fm/c) to about 8 fm/c the ratio of the scalar quark condensate remains
far below unity, which implies that the decay of strings in the
hadronic environment is modified substantially in the hot and dense
medium.

The white borderlines in Fig. \ref{fig2} separate the space-time
regions of deconfined matter to hadronic matter.  It is clearly seen
that although the chiral symmetry is approximately restored in the
full overlap phase from 2.8 to 6 fm/c some region is occupied
by deconfined partons in the PHSD. In these regions, however, an
enhanced production of strangeness should not occur since the
Schwinger mechanism (\ref{schwinger}) no longer applies due to a
vanishing string tension and a transformation of energy and momentum to
massive partonic degrees-of-freedom.

 \begin{figure}[thb]
\includegraphics[width=0.37\textwidth]{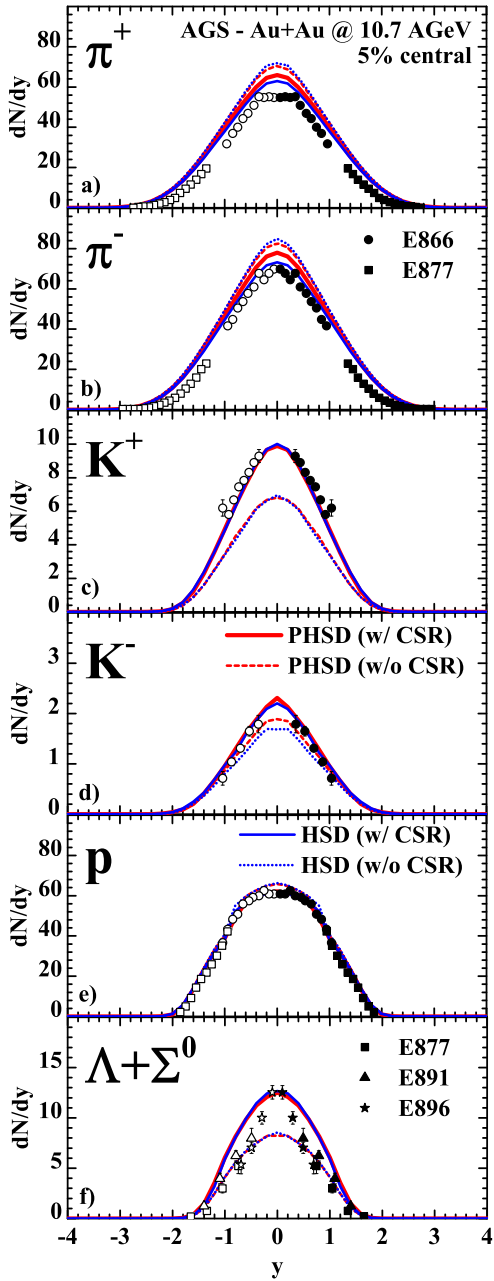}
\caption{The rapidity distribution of pions,  kaons, protons and
$(\Lambda+\Sigma^0)$'s for 5\% central  Au+Au collisions at 10.7  A GeV in
comparison to the experimental data from Ref. \cite{exp1}. The solid
(red) lines show the results from PHSD (including CSR) while the
blue solid lines result from HSD (including CSR) without partonic
degrees-of-freedom. The dashed (red) line reflects the PHSD results
without CSR while the dashed blue line results from HSD without
CSR.}
 \label{fig3}
\end{figure}

  \begin{figure}[thb]
\includegraphics[width=0.37\textwidth]{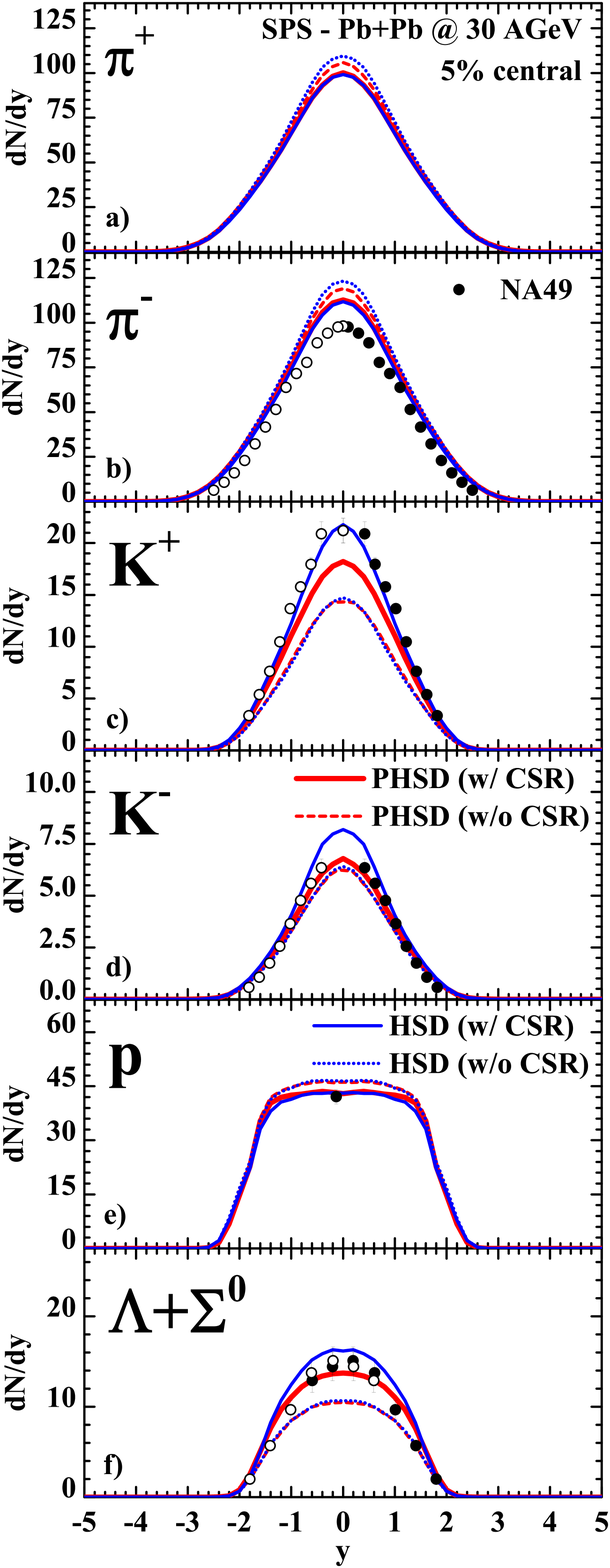}
\caption{The rapidity distribution of pions,  kaons, protons and $(\Lambda+\Sigma^0)$'s
for 5\% central  Au+Au collisions at 30  A GeV in comparison to the experimental data from Ref. \cite{exp2}.
The coding of the lines is the same as in Fig. \ref{fig3}.}
 \label{fig4}
\end{figure}

 \begin{figure}[thb]
\includegraphics[width=0.37\textwidth]{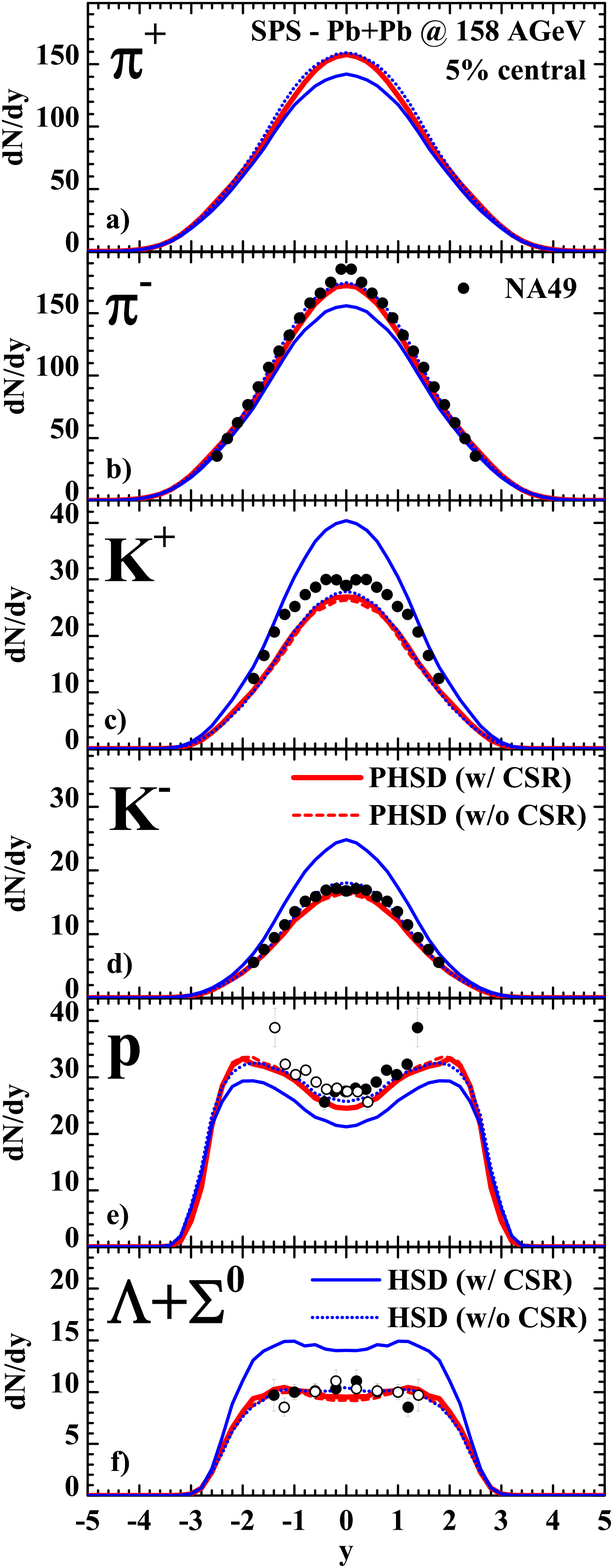}
\caption{The rapidity distribution of pions,  kaons, protons and $(\Lambda+\Sigma^0)$'s
for 5\% central  Au+Au collisions at 158  A GeV in comparison to the experimental data from Ref. \cite{exp3}.
The coding of the lines is the same as in Fig. \ref{fig3}.}
 \label{fig5}
\end{figure}

After these more quantitative illustrations we continue with actual
observables from heavy-ion reactions at different energies. We will consider
the following four scenarios:
\begin{itemize}
\item{The default HSD calculations without any CSR (and deconfinement transition) and a threshold in the local energy
density of $<$ 0.5 GeV/fm$^3$ for elastic and inelastic reactions of
formed hadrons. For energy densities above  0.5 GeV/fm$^3$ a free
streaming of the particles is assumed.}
\item{HSD calculations with the modified string decay (describing CSR) for all local energy
densities, however, keeping the free streaming of hadrons above 0.5
GeV/fm$^3$. Also in this scenario there is no deconfinement
transition.}
\item{The default PHSD calculations without any CSR, however, a crossover transition to the deconfined
phase above the threshold in the local energy density of 0.5
GeV/fm$^3$.}
\item{PHSD calculations with the modified string decay (describing CSR) for
energy densties below 0.5 GeV/fm$^3$ and  a crossover transition to
the deconfined phase above the threshold in the local energy density
of 0.5 GeV/fm$^3$.}
\end{itemize}

The results for the rapidity distributions of pions, kaons, protons
and $(\Lambda+\Sigma^0)$'s in the different limits are shown in
Figs. \ref{fig3} to \ref{fig5} for 5\% central  Au+Au collisions at
10.7, 30 and 158 A GeV in comparison to the experimental data from
Refs. \cite{exp1,exp2,exp3}. Here the solid (red) lines show the
results from PHSD (including CSR) while the blue solid lines result
from HSD (including CSR) without partonic degrees-of-freedom. The
dashed (red) line reflects the PHSD results without CSR while the
dashed blue line results from HSD without CSR. As noted before, the
HSD results (without CSR) for these hadrons are essentially the same
than in the previous studies \cite{Jgeiss,Brat04,Weber} and severely
underestimate the $K^+$ and $\Lambda$ production at the lower
energies while overproducing pions. Furthermore, the partonic
degrees-of-freedom in PHSD -- including a phase transition to the
QGP in case of sufficient local energy density -- do not change the
rapdity distributions (red dashed lines) compared to HSD at these
energies. Actually this rough equivalence also holds for PHSD and
HSD when including CSR (red and dashed solid lines) at 10.7 A GeV,
however, in this case the $K^\pm$ and $(\Lambda+\Sigma^0)$
distributions are better in line with the experimental observations.

We note that our 'HSD' results are up to 10\% lower than those from
Refs. \cite{Brat04,Weber}. As discussed in section II.C. this is
essentially due to a reduction of the critical energy density from 1
GeV/fm$^3$ to 0.5 GeV/fm$^3$ according to more recent lattice QCD
results on the critical temperate $T_c$ for 2+1 flavors
\cite{lqcd0}. This reduces the hadronic scattering rate since
hadrons in local cells with energy densities above 0.5 GeV/fm$^3$ are
freely streaming.

Although the  $\pi^\pm$ multiplicities are still overestimated by
the PHSD calculations (including CSR) at 10.7 and 30 A GeV a
striking improvement is obtained with respect to the strangeness
production at these energies for central Au+Au collisions. Since
this result emerges without any fine-tuning of the nuclear equation
of state (parameter-sets NL1, ML2, NL3) we attribute the strangeness
enhancement seen in these reactions to the approximate restoration
of chiral symmetry at high baryon density. This interpretation is in
contrast to the early expectation in Refs.
\cite{rafelski,GG99,GGS11} that the enhancement of strangeness
should be attributed to the formation of deconfined matter.
 \begin{figure}[thb]
\includegraphics[width=0.4\textwidth]{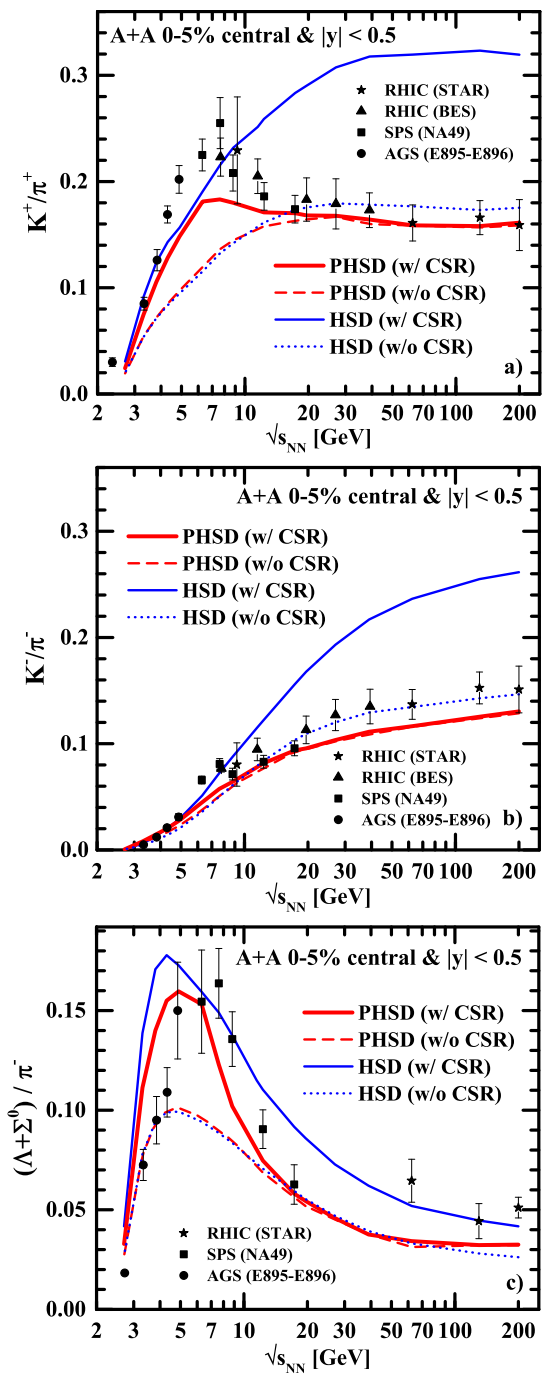}
\caption{The ratios $K^+/\pi^+$, $K^-/\pi^-$ (a) and $\Lambda/\pi^-$
(b) at midrapidity from 5\% central Au+Au collisions as a function
of the invariant energy $\sqrt{s_{NN}}$ up to the top RHIC energy in
comparison to the experimental data from \cite{exp4}. The coding of
the lines is the same as in Fig. \ref{fig3}.}
 \label{fig6}
\end{figure}

In order to summarize our findings we show the ratios $K^+/\pi^+$,
$K^-/\pi^-$ and $(\Lambda+\Sigma^0)/\pi^-$ at midrapidity from 5\%
central Au+Au collisions in Fig. \ref{fig6} as a function of the
invariant energy $\sqrt{s_{NN}}$ up to the top RHIC energy in
comparison to the experimental data available \cite{exp4}.    As
before the solid (red) lines show the results from PHSD (including
CSR) while the blue solid lines result from HSD (including CSR)
without partonic degrees-of-freedom. The dashed (red) line reflects
the PHSD results without CSR while the dashed blue line results from
HSD without CSR. It is clearly seen from Fig. \ref{fig6} that the
results from HSD and PHSD merge for $\sqrt{s_{NN}} <$ 6 GeV and fail
to describe the data in the conventional scenario without
incorporating the CSR as described in Section II (and been found
earlier in Refs. \cite{Brat04,Weber}). Especially the rise of the
$K^+/\pi^+$ ratio at low bombarding follows closely the experimental
excitation function when incorporating  'chiral symmetry
restoration'. However, the drop in this ratio again is due to
deconfinement since there is no longer any string decay in a
hadronic medium at higher bombarding energies. This is clearly seen
in the case of HSD (with CSR) which overshoots the data
substantially at high bombarding energy.

Accordingly the experimental 'horn' in the excitation function is
caused by chiral symmetry restoration but also deconfinement is
essential to observe a maximum in the $K^+/\pi^+$  ratio. We mention
that the maximum in the $K^+/\pi^+$  ratio is not so pronounced in
the PHSD calculations as in the data since the pion production is
still overestimated by the PHSD. We speculate that this
overestimation might be due to the complex pion dynamics in the
nuclear medium where the pion separates into a 'pion' and
'$\Delta$-hole' branch \cite{Ehe}.

\section{Summary}\label{summary}

We have studied effects from chiral symmetry restoration (CSR) in
relativistic heavy-ion collisions in the energy range from 4 to 158
A GeV by using the Parton-Hadron-String Dynamics (PHSD) approach
\cite{PHSDrhic} that has been extended essentially in the hadronic
phase to also describe CSR apart from a deconfinement transition to
dynamical quarks, antiquarks and gluons in the QGP.

We have assumed that effects of chiral symmetry restoration for the
scalar quark condensate $<q {\bar q}>$ in the hadronic medium show
up in the Schwinger formula (\ref{schwinger}) for the $s/u$ ratio
when the string decays in a dense or hot hadronic medium. The
evaluation of the scalar quark condensate has been based on Eq.
(\ref{condens2}) where dominantly the quantity $\Sigma_\pi \approx$
45 MeV enters as well as the scalar nucleon density $\rho_S$ that
drives CSR at low temperatures of the system.
Although the value of $\Sigma_\pi$ might be slightly different
\cite{Meissner,Sternbeck} we have kept this conservative value for
our studies. The computation of the scalar baryon density
$\rho_S({\bf r};t)$ has been based on the nonlinear $\sigma-\omega$
model for nuclear matter \cite{Boguta,Lang} and in particular on the
gap equation (\ref{gap}) for each cell of the space-time grid
employing different parameter-sets. These equations complete the
description of the extended PHSD approach. We note in passing that
we just have used 'default values' for the couplings and low-energy
constants and not attempted to perform any fitting.

When comparing the results from the extended PHSD approach for the
ratios $K^+/\pi^+$, $K^-/\pi^-$ and $(\Lambda+\Sigma^0)/\pi^-$ from
the different scenarios we find from Fig. \ref{fig6} that the
results from HSD and PHSD merge for $\sqrt{s_{NN}} <$ 6 GeV and fail
to describe the data in the conventional scenario without
incorporating the CSR as described in Section II. Especially the
rise of the $K^+/\pi^+$ ratio at low bombarding follows closely the
experimental excitation function when including 'chiral symmetry
restoration' in the string decay. However, the drop in this ratio
again is due to 'deconfinement' since there is no longer any
hadronic string decay in a partonic medium at higher bombarding
energies. This is clearly seen in the case of HSD (with CSR) which
overshoots the data substantially at high bombarding energy.
Accordingly the experimental 'horn' in the excitation function is
caused by chiral symmetry restoration but also deconfinement is
essential to observe a maximum in the $K^+/\pi^+$ and
$\Lambda/\pi^-$ ratios. The maximum in the PHSD calculations is not
very pronounced since the pion abundance is still overestimated in
this energy range. Our interpretation differs from the early
expectation in Refs. \cite{rafelski,GG99,GGS11} that the enhancement
of strangeness should be attributed to the formation of deconfined
matter. Our studies thus support the conjecture of 'quarkyonic
matter' \cite{Rob1} out-off equilibrium in central heavy-ion
collisions from about 5 to 40 A GeV.

We finally recall that a fully consistent approach has to include
not only the chiral effects on the hadron production but also the
chiral partners of the $0^-$ and $1^-$ mesons, i.e. the (broad)
scalar $0^+$ and axial vector $1^+$ mesons with their spectral
functions. Also on the baryonic side the chiral partners to the
lightest baryons have to be incorporated in the transport approach
with dynamical spectral functions that become identical in the
chiral limit. Since the PHSD approach is suited for this task due to
its off-shell nature for all degrees-of-freedom we delay a
selfconsistent treatment of the chiral dynamics to a future
more elaborate study. \\

\section*{Acknowledgements}
The authors acknowledge inspiring discussions with J. Aichelin, D.
Cabrera, V. P. Konchakovski, O. Linnyk, R. Stock and V. D. Toneev.
This work was in part supported by BMBF.  The computational
resources have been provided by the LOEWE-CSC as well as the SKYLLA
cluster at the Univ. of Giessen.


\begin{thebibliography}{99}
\bibitem{lQCD} Y. Aoki {\it et al.}, Phys. Lett. B {\bf 643}, 46 (2006).
\bibitem{lqcd0}  S. Borsanyi {\it et al.},
JHEP {\bf 1009}, 073 (2010); JHEP {\bf 1011}, 077 (2010); JHEP
{\bf 1208}, 126 (2012).
\bibitem{LQCDx} S. Borsanyi {\it et al.}, Phys. Lett B {\bf 730}, 99
(2014); Phys. Rev. D {\bf 92}, 014505 (2015).

\bibitem{Peter} P. Petreczky [HotQCD Collaboration], PoS LATTICE
{ \bf 2012}, 069 (2012); AIP Conf. Proc. {\bf 1520}, 103 (2013).

\bibitem{Lat1} H.-T. Ding, F. Karsch, S. Mukherjee, arXiv:1504.05274

\bibitem{Lat2} A. Bazavov {\it et al.}, Phys. Rev. D {\bf 90},
094503 (2014).

\bibitem{QM2014}
Proceedings of ’Quark Matter-2014’, Nucl. Phys. A {\bf 931}, 1 (2014).

%
\bibitem{CBMbook} P. Senger {\it  et al.}, Lect. Notes Phys. {\bf 814}, 681 (2011).

\bibitem{Ch1}
C. S. Fischer, J. Luecker, and C. Welzbacher, {Phys. Rev.} D {\bf 90}, 034022 (2014).
\bibitem{Ch2}  C. S. Fischer, L.  Fister, J. Luecker, and  J. M. Pawlowski,
 Phys. Lett. B {\bf 732}, 273 (2014).
\bibitem{BJS} T. K. Herbst, J. M. Pawlowski, and B.-J. Schaefer,
Phys. Rev. D {\bf 88}, 014007 (2013).

\bibitem{Weise}
    U. Vogl and W. Weise, Prog. Part. Nucl. Phys. {\bf 27}, 195 (1991).
\bibitem{Birse}
    M. C. Birse, J. Phys. G {\bf 20}, 1537 (1994).

\bibitem{Kochr}
    V. Koch, Int. J. Mod. Phys. E  {\bf 6}, 203 (1997).
\bibitem{Zahed}
        J. V. Steele, H. Yamagishi, and I. Zahed,
        Phys. Lett. B  {\bf 384}, 255 (1996);  Phys. Rev. D  {\bf 56}, 5605 (1997).
\bibitem{GEB}
    G. E. Brown, Prog. Theor. Phys.  {\bf 91}, 85 (1987).
\bibitem{GEB1}
    G. E. Brown, C. M. Ko, Z. G. Wu, and L. H. Xia,
    Phys. Rev. C  {\bf 43}, 1881 (1991).
\bibitem{Koch}
    V. Koch and G. E. Brown, Nucl. Phys. A  {\bf 560}, 345 (1993).
\bibitem{Stoecker}
    H. St\"ocker and W. Greiner, Phys. Rep.  {\bf 137}, 277 (1986).
\bibitem{Bertsch}
    G. F. Bertsch and S. Das Gupta, Phys. Rep.  {\bf 160}, 189 (1988).
\bibitem{Cass90}
    W. Cassing, V. Metag, U. Mosel, and K. Niita,
       Phys. Rep.  {\bf 188}, 363 (1990).
\bibitem{Hart}
C. Hartnack {\it et al.},
Eur. Phys. J. A {\bf 1}, 151  (1998).

\bibitem{HSD}
      W. Cassing and E. L. Bratkovskaya, Phys. Rep. {\bf 308}, 65 (1999).

\bibitem{Koreview}
    C. M. Ko and G. Q. Li, J. Phys. G  {\bf 22}, 1673 (1996).

\bibitem{UrQMD1}
    S. A.~Bass {\it et al.},
M.~Belkacem, M.~Bleicher, M.~Brandstetter, L.~Bravina,
    C.~Ernst, L.~Gerland, M.~Hofmann, S.~Hofmann, J.~Konopka, G.~Mao,
    L.~Neise, S.~Soff, C.~Spieles, H.~Weber, L. A.~Winckelmann,
    H.~St\"ocker, W.~Greiner, Ch.~Hartnack, J.~Aichelin, and N.~Amelin,
    Prog. Part. Nucl. Phys. {\bf 42}, 279 (1998).
\bibitem{Brat2012} C. Hartnack,
H. Oeschler, Y. Leifels, E. L. Bratkovskaya, and J. Aichelin, Phys.
Rept. {\bf 510}, 119 (2012).

\bibitem{Marty} R. Marty and J. Aichelin, Phys. Rev. C {\bf 87}, 034912 (2013); R. Marty {\it et al.},
Phys. Rev. C {\bf 92}, 015201 (2015).
\bibitem{Bot}
    W. Botermans and R. Malfliet, Phys. Rep.  {\bf 198}, 115 (1990).
\bibitem{Mal}
    R. Malfliet, Prog. Part. Nucl. Phys.  {\bf 21}, 207 (1988).
\bibitem{T3}
    A. Faessler, Prog. Part. Nucl. Phys.  {\bf 30}, 229 (1993).


\bibitem{Laura} L. Tol\'os, A.  Ramos, and A. Polls, Phys. Rev. C
{\bf 65}, 054907   (2002).


\bibitem{EX4} W. Cassing, L. Tol\'os, E. L. Bratkovskaya, and A. Ramos,
Nucl. Phys. A {\bf 727}, 59 (2003).

\bibitem{rafelski} J. Rafelski and B. M\"uller, Phys. Rev. Lett.
{\bf 48}, 1066 (1982).

\bibitem{stock} R. Stock, J. Phys. G {\bf 28}, 1517 (2002).

\bibitem{GG99} M. Gazdzicki and M. I. Gorenstein, Acta Phys. Polon.
B {\bf 30}, 2705 (1999).

\bibitem{GGS11} M. Gazdzicki, M. Gorenstein and P. Seyboth,
Acta Phys. Polon. B {\bf 42}, 307 (2011).

\bibitem{Jgeiss}  J. Geiss, { W. Cassing}, and C. Greiner,
Nucl. Phys. A {\bf 644}, 107 (1998).

\bibitem{Brat04} E. L. Bratkovskaya {\it et al.},
Phys. Rev. C {\bf 69}, 054907 (2004).

\bibitem{Weber} H. Weber, E. L. Bratkovskaya, W. Cassing, and H. St\"ocker,
Phys. Rev. C {\bf 67}, 014904 (2003).


\bibitem{Petersen} H. Petersen et al., Phys. Rev. C {\bf 78}, 044901
(2008).

\bibitem{hydro2} Yu. B. Ivanov, V. N. Russkikh, and V.D. Toneev,
 Phys. Rev. C {\bf 73}, 044904  (2006);
Yu. B. Ivanov and D. Blaschke,  Phys. Rev. C {\bf 92}, 024916 (2015).



\bibitem{ABS06} A. Andronic, P. Braun-Munzinger and J. Stachel, Nucl.
Phys. A {\bf 772}, 167 (2006); Phys. Lett. B {\bf 673}, 142 (2009).
\bibitem{Brat00}  E. L. Bratkovskaya, { W. Cassing}, C. Greiner,
M. Effenberger, U. Mosel, and A. Sibirtsev,
Nucl. Phys. A {\bf 675}, 661 (2000).

%
\bibitem{Vitalii}  V. Ozvenchuk, O. Linnyk, M.I. Gorenstein, E. L.
Bratkovskaya, and  W. Cassing,
Phys. Rev. C {\bf 87},  024901 (2013)

\bibitem{Rob1} L. McLerran and R. D. Pisarski,  Nucl. Phys. A {\bf 796},83 (2007).


\bibitem{Rob2} A. Andronic {\it et al.},
 Nucl. Phys. A {\bf 837}, 65 (2010).

\bibitem{sasaki} K. Fukushimi and C. Sasaki, Prog. Part. Nucl. Phys.
{\bf 72}, 99 (2013).

\bibitem{Nambu1961} Y. Nambu and G. Jona-Lasinio, Phys. Rev. {\bf 122}, 345 (1961).

\bibitem{Klevansky1992} S. P. Klevansky, Rev. Mod. Phys. {\bf 64}, 649 (1992).

\bibitem{NJL}
R. Marty {\it et al.},
Phys. Rev. C {\bf 88}, 045204 (2013).

\bibitem{Juan} J. M. Torres-Rincon, B. Sintes, and J. Aichelin,
 Phys. Rev. C {\bf 91}, 065206 (2015).




\bibitem{PHSD}
W. Cassing and E.L. Bratkovskaya, Nucl. Phys. A {\bf 831}, 215 (2009).

\bibitem{PHSDrhic}
  E.~L.~Bratkovskaya, W.~Cassing, V.~P.~Konchakovski and O.~Linnyk,
  Nucl.\ Phys.\ A {\bf 856}, 162 (2011).
%

\bibitem{Volo} V. P. Konchakovski {\it et al.},
Phys. Rev. C {\bf 85},  044922 (2012);  Phys. Rev. C {\bf 85}, 011902 (2012).

\bibitem{Linnyk} O. Linnyk {\it et al.},  Phys. Rev. C {\bf 89},   034908 (2014);
Phys. Rev. C {\bf 88},  034904 (2013); Phys. Rev. C {\bf 87},   014905 (2013);
Phys. Rev. C {\bf 85},  024910 (2012); Phys. Rev. C {\bf 84}, 054917 (2011); Nucl. Phys. A {\bf 855}, 273 (2011).

\bibitem{TSong}  T. Song, H. Berrehrah, D. Cabrera, J. M.
Torres-Rincon, L. Tolos, { W. Cassing}, and E. L.  Bratkovskaya,
Phys. Rev. C {\bf 92}, 014910 (2015).

\bibitem{Volo14}  V. P. Konchakovski, { W. Cassing}, Yu. B. Ivanov,
and V.D. Toneev,
Phys. Rev. C {\bf 90}, 014903 (2014).

\bibitem{BratPRL} E. L. Bratkovskaya, S. Soff, H. St\"ocker, M. van Leeuwen,
W. Cassing,
 Phys. Rev. Lett. {\bf 92}, 032302 (2004).

%
\bibitem{Kadanoff1}
L. P. Kadanoff and G. Baym,  {\it Quantum Statistical Mechanics},
Benjamin, New York, 1962.
%
\bibitem{Kadanoff2}
S. Juchem, W. Cassing,  and C. Greiner, { Phys. Rev.} D {\bf 69}, 025006 (2004);
{ Nucl. Phys.} A {\bf 743}, 92 (2004).

\bibitem{Cassing:2008nn}
  W.~Cassing,
  Eur.\ Phys.\ J.\ ST {\bf 168}, 3 (2009); Nucl. Phys. A {\bf 795}, 70 (2007).
\bibitem{Vitaly2}
V. Ozvenchuk, O. Linnyk, M.I. Gorenstein, E.L. Bratkovskaya and  W. Cassing,
 Phys.Rev. C {\bf 87},  064903 (2013).

\bibitem{Ca13}
W. Cassing, O. Linnyk, T. Steinert, and V. Ozvenchuk,
 Phys. Rev. Lett. {\bf 110},  182301 (2013); T. Steinert and W. Cassing,  Phys. Rev. C {\bf 89},  035203 (2014).



\bibitem{FRITIOF}     B. Nilsson-Almqvist and E. Stenlund, Comp. Phys. Comm.
{\bf 43}, 387     (1987);
B. Andersson, G. Gustafson, and H. Pi, Z. Phys. C {\bf 57}, 485 (1993).


\bibitem{Sjostrand:2006za}
  T.~Sjostrand, S.~Mrenna and P.~Z.~Skands,
  JHEP {\bf 0605}, 026 (2006).

%
%
\bibitem{PHSDLHC} V. Konchakovski, W. Cassing and V. D. Toneev,
J. Phys. G {\bf 42}, 055106 (2015); J. Phys. G {\bf 41}, 105004 (2014).
%
\bibitem{Schwinger}
    J. Schwinger, Phys. Rev. {\bf 83}, 664 (1951).

 \bibitem{Ca02} W. Cassing,
Nucl. Phys. A {\bf 700}, 618 (2002).

\bibitem{Song1}  C. H. Li and C. M. Ko, Nucl. Phys. A {\bf 712},110 (2002).
\bibitem{Song2}  F. Li, L. W. Chen and C. M. Ko, Phys. Rev. C {\bf 85}, 064902 (2012).
\bibitem{Song3} V. Flaminio, W. G. Moorhead, D. R. O. Morrison, and N. Rivoire, {\it CERN-HERA-83-02}.


\bibitem{Meissner} M. Hoferichter, J. Ruiz de Elvira, B. Kubis, and U.-G.
Meissner, arXiv:1506.04142

\bibitem{Sternbeck} A. Sternbeck, to be published.


\bibitem{Toneev98}
    B. Friman, W.  N\"orenberg and V. D. Toneev,
    Eur. Phys. J. A {\bf 3}, 165 (1998).

\bibitem{GOR}
    M. Bando, T. Kugo and K. Yamawaki, Phys. Rep. {\bf 164}, 217 (1988).
\bibitem{Cohen}
    T. D. Cohen, R. J. Furnstahl, D. K. Griegel, and X. Jin,
    Prog. Part. Nucl. Phys. {\bf 35}, 221 (1995).
\bibitem{Boguta} J. Boguta and A. R. Bodmer,  Nucl. Phys. A {\bf 292}, 413 (1977).

\bibitem{Lang} A. Lang et al., Z. Phys. A
{\bf 340}, 287 (1991).


\bibitem{DBR} F. de Jong and  R. Malfliet,  Phys. Rev. C {\bf 44}, 998 (1991).

\bibitem{Dejong89}
F.~de~Jong, B.~ter Haar, and R.~Malfliet, Phys. Lett. B {\bf 220}, 485 (1989).

\bibitem{CBJ00} W. Cassing, E. L. Bratkovskaya, and S. Juchem,
Nucl. Phys. A {\bf 674}, 249 (2000).

\bibitem{exp1}
    J.~Stachel,
    Nucl.\ Phys.\ A {\bf 610}, 509C (1996);
    S.~Albergo {\it et al.},
    Phys.\ Rev.\ Lett.\  {\bf 88}, 062301 (2002).
\bibitem{exp2}
    C.~Alt {\it et al.} [NA49 Collaboration],
    Phys.\ Rev.\ C {\bf 77}, 024903 (2008);
    C.~Alt {\it et al.} [NA49 Collaboration],
    Phys.\ Rev.\ C {\bf 73}, 044910 (2006);
    C.~Alt {\it et al.} [NA49 Collaboration],
    Phys.\ Rev.\ C {\bf 78}, 034918 (2008).
\bibitem{exp3}
    S.~V.~Afanasiev {\it et al.} [NA49 Collaboration],
    Phys.\ Rev.\ C {\bf 66}, 054902 (2002);
    T.~Anticic {\it et al.} [NA49 Collaboration],
    Phys.\ Rev.\ C {\bf 83}, 014901 (2011);
    T.~Anticic {\it et al.} [NA49 Collaboration],
    Phys.\ Rev.\ Lett.\  {\bf 93}, 022302 (2004).
\bibitem{exp4}
    B.~I.~Abelev {\it et al.} [STAR Collaboration],
    Phys.\ Rev.\ C {\bf 81}, 024911 (2010);
    M.~M.~Aggarwal {\it et al.} [STAR Collaboration],
    Phys.\ Rev.\ C {\bf 83}, 024901 (2011).

\bibitem{Ehe} W. Ehehalt, W. Cassing, A. Engel, U. Mosel, and G. Wolf,
Phys. Lett. B {\bf 298}, 31 (1993).



\end{thebibliography}
\end{document}